\documentclass[aps, prb , 10pt ,a4paper,twocolumn,floatfix,showpacs,showkeys,amsmath,amssymb,altaffilletter]{revtex4-1}

\usepackage{graphicx}
\usepackage{epstopdf,amsmath,amssymb, booktabs}

\begin{document}

\title{Organic solar cell efficiencies under the aspect of reduced surface recombination velocities}

\author{A. Wagenpfahl}
\author{C. Deibel}
\affiliation{Experimental Physics VI, Julius-Maximilians-University of W{\"u}rzburg, D-97074 W{\"u}rzburg.}
\author{V. Dyakonov}
\affiliation{Experimental Physics VI, Julius-Maximilians-University of W{\"u}rzburg, D-97074 W{\"u}rzburg and\\ Functional Materials for Energy Technology, Bavarian Centre for Applied Energy Research (ZAE Bayern), D-97074 W{\"u}rzburg.}

\date{\today}

\begin{abstract}

The charge carrier mobility is a key parameter for the organic bulk heterojunction solar cell efficiency. It was recently shown that the interplay charge carrier transport and recombination, both depending on electron and hole mobilities, leads to a point of maximum power conversion efficiency at a finite mobility. Changes of bulk and surface recombination rate, however, can strongly influence this behavior. These processes were previously not considered adequately, as surface recombination velocities of infinity were implicitly assumed or bulk recombination parameters not discussed in detail. In this manuscript, using a macroscopic effective medium simulation, we consider how a reduced bulk recombination process in combination with finite surface recombination velocities affect the power conversion efficiency. Instead of a maximum efficiency at a specific charge carrier mobility, we show that with realistic assumptions and passivated surfaces the efficiency is increased further, saturating only at higher mobilities.  Thus, a mobility optimisation is more important for the solar cell performance then previously shown.

\end{abstract}



\maketitle
\newpage

\section{Introduction}

High power conversion efficiencies of organic solar cells are one key criterium for a use of organic semiconductors in industrial applications. Until now, by improving process parameters, bulk heterojunction solar cell efficiencies of up to 5\% are reached by several groups using well known materials like poly(3-hexylthiophene) (P3HT) and phenyl-C61-butyric acid methyl ester (PCBM) blends \cite{green_x2008,peet2007}. Recently, by introducing new organic semiconductors providing advantageous energy levels and charge carrier mobilities, verified efficiencies of up to 6\% and internal quantum yields of almost 100\% could be achieved \cite{park2009}. These examples show how important it is to design optimized organic semiconductor for solar cell applications. In order to predict the influence of charge carrier mobilities, numerical calculations have been performed, showing power conversion maxima at a certain mobility equal for electrons and holes \cite{deibel2008a, mandoc2007}.Further macroscopic simulations based on the detailed balance framework confirmed this prediction, but pointed out that a finite surface recombination at the electrodes can avoid a lowering of the power conversion efficiency at high charge carrier mobilities \cite{kirchartz2008}. As simplification, all those calculation either implicitly assumed an infinite surface recombination rate for the contacts or do not discuss the influence of such details like the combined effects of finite surface recombination rates and a limited  bulk, e.g. bimolecular, recombination.\\

In this paper, we present how charge carrier mobilities and surface recombination processes of electrons and holes influence the power conversion efficiencies of an organic solar cell. We do this by performing numerical simulations, also taking a physically reasonable reduced and capped bimolecular Langevin-like recombination into account \cite{deibel2009}.

\section{Simulation}

To calculate current--voltage characteristics of organic solar cells, a differential equation system of discretized Poisson and continuity equations for electrons and holes has to be solved as already shown in \cite{deibel2008a, hausermann2009, koster2005b}.\\

For the active material the effective medium approach is used, based on a blend of organic semiconducting donor (polymer) and acceptor (fullerene) molecules under consideration of reachable donor acceptor interfaces within the diffusion length of charge carriers everywhere inside the active region. Due to the energetic advantage excitions created e.g. by incident light on one molecule tend to separate to a coulombicly bound charge transfer state on a femto second timescale \cite{sariciftci1992}. This leads to the model of an effective electrical band gap $E_g$ between the acceptor LUMO (lowest unoccumpied molecular orbital) and the donor HOMO (highest occupied molecular orbital) level, being able to describe the electrical properties of the whole device in terms an ambipolar effective medium with a band-like transport. For the numerical description, also input parameters such as the thickness $L$, the effective charge carrier densities $N_c$, $N_v$ for HOMO  and LUMO bands, the charge carrier mobilities for electrons $\mu_n$ and holes $\mu_p$ and the relative dielectric permeability $\epsilon_r$ are required. Also the electron charge $q$, the vacuum permeability $\epsilon_0$ and the Boltzmann constant $k_B$ are used.\\

On the active layer a predominantly electron conducting cathode and a predominantly hole conducting anode are added. Discontinuities of electrode work functions to the corresponding active material bands are considered by injection barriers $\Phi_a$, $\Phi_c$. Furthermore, both contacts were assumed to possess finite surface recombination velocities. DDue to these contacts, the active material next to the cathode (anode) possesses a much higher electron (hole) charge carrier density, therefore, called the majority and a lower minority hole (electron) density. Consequently, within this paper, we denote the surface recombination, electrons crossing the cathode and holes crossing the anode interface, as the majority surface recombination $S_{Maj}$, and the vice versa, holes crossing the cathode and electrons crossing the anode interface, to be theminority surface recombination $S_{Min}$. Reduced (finite) majority surface recombination velocities describe charges taking time to overcome injection or extraction barriers at the interface. For minority charges, a reduced surface recombination velocity results in a lower current of minority charges diffusing into the electrode, e.g. caused by mirror charge effects at the metallic electrode. In general real interfaces do not possess infinite recombination velocities. At the electrode minority charges recombine instantaneously with the present majority charges. Thus, a reduced minority surface recombination passivates the surface.\\

At a given temperature $T$, we are now able to calculate the potential energy levels of $E_{LUMO}$, $E_{HOMO}$ and the corresponding quasi-Fermi levels of electrons $E_{Fn}$ and holes $E_{Fp}$ as well as the charge carrier density distribution inside the solar cell for electrons $n$ and holes $p$. Energy levels and charge carrier densities are linked by the Boltzmann distribution, e.g. for electrons $n=N_c \exp\left(- q (E_{LUMO} - E_{Fn}) / k_b T\right)$.\\

Illumination of a solar cell can be described by adding a spatially constant charge carrier excitation $G$. This is opposed by a bimolecular, non-geminate recombination rate $R$, proven in disordered systems to accord to the theory of Langevin, describing recombination of charges residing on different phases of the effective medium \cite{langevin1903, pope1999}. Recently, it was found that the experimental determined bimolecular recombination rate is reduced in comparison to Langevins' theory \cite{deibel2009, deibel2008b, juska2008, pivrikas2005, shuttle2008}. We could explain the temperature dependence of this reduction factor by considering the spatial charge carrier distribution gradients inside a device \cite{deibel2009}. Nevertheless, the origin of the remaining constant reduction factor of $\zeta = 0.1$ is unknown so far. With this factor, the bimolecular recombination rate reads
\begin{equation}
	R=\zeta \gamma \left( n p - n_i^2 \right),
	\label{eqn:langevin}
\end{equation}
with
\begin{equation}
	\gamma=\frac{q}{\epsilon_0 \epsilon_r}\left( \mu_n + \mu_p \right)
	\label{eqn:prefactor}
\end{equation}\\
and the intrinsic charge carrier concentration
\begin{equation}
	n_i=\sqrt{N_c  N_v} \exp \left( -q E_g / 2 k_b T \right).
\end{equation}

However, the reduced recombination rate is proved to be valid in low mobility materials only (\mbox{$\lesssim1$cm$^2$/Vs}) \cite{pope1999}. In terms of Eqn. (\ref{eqn:prefactor}), faster charges would have a higher, unlimited recombination probability. Even so, as this assumption only holds for low mobilities, at high mobilities the bimolecular recombination is mobility independent, leading to a constant Langevin prefactor $\gamma$. This can be compared with the mobility independent recombination rates of high mobility inorganic systems such as the compound semiconductor Cu(In,Ga)Se$_2$.\\

A possible but crude approximation simply cuts off the Langevin prefactor $\gamma$ at a certain point, now defined as
\begin{equation} 
 \gamma =\begin{cases} 
	\frac{q}{\epsilon_0 \epsilon_r}\left( \mu_n + \mu_p \right) & \text{if} ~~ \frac{q}{\epsilon_0 \epsilon_r}\left( \mu_n + \mu_p \right) \leq \gamma_{crit}\\
	\gamma_{crit} & \text{if} ~~ \frac{q}{\epsilon_0 \epsilon_r}\left( \mu_n + \mu_p \right) > \gamma_{crit}.
 \end{cases}
 \label{eqn:cutoff}
\end{equation}

Since $\gamma_{crit}$ is a physically reasonable but not directly accessible parameter, we convert it to a critical mobility $\mu_{crit} = 2\frac{\epsilon_0 \epsilon_r}{q}\gamma_{crit} $, assuming identical values for electrons and holes. In the following, we denote the loss according to Eqn. (\ref{eqn:cutoff}) as capped Langevin recombination. A general overview of the underlying processes can be found in \cite{brabec2008book}.\\

\begin{table}
	\centering
		\begin{tabular}{llll}		
			\toprule
			\hline		
			parameter & symbol & value & unit \\			
			\midrule
			\hline
			temperature & $T$ & 300 & K\\
			effective band gap & $E_g$ & 1.05 & eV\\
			relative dielectric constant & $\epsilon_r$ & $4.0$ & \\
			active layer thickness & $L$ & 100 & nm\\
			effective density of states & $N_c,~N_v$ & $1.0 \times 10^{26}$ & 1/m$^{3}$\\
			generation rate & $G$ & $9.0 \times 10^{27}$ & 1/m$^3$s\\
			injection barriers & $\Phi_a,~\Phi_c$ & $0.1$ & eV\\
			Langevin reduction factor & $\zeta$ & 0.1\\
			\bottomrule
			\hline
		\end{tabular}
	\caption{Parameters used for simulation.}
	\label{tab:param}
\end{table}The presented calculation parameters listed in Table \ref{tab:param}, used for all presented calculations, were acquired by fitting the current--voltage characteristics of a representative 5\% P3HT:PCBM solar cell (not shown) under assumption of infinite surface recombination velocities. The charge carrier mobilities where determined experimentally \cite{baumann2008}. 

\section{Results}

\subsection{Variation of charge carrier mobility}

\begin{figure}
	\centering\includegraphics[width=9.0cm]{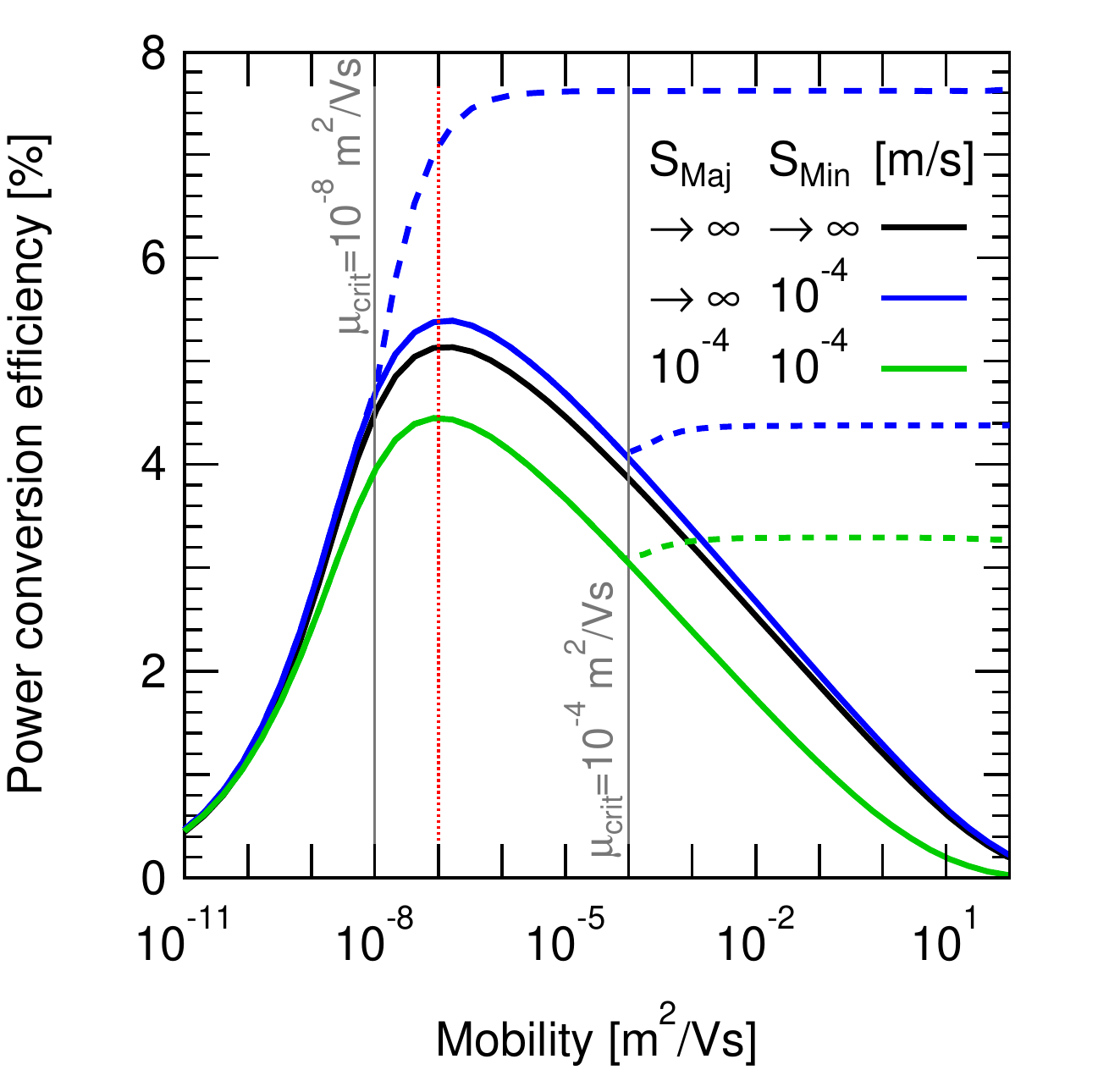}
	\caption{(Color online) Power conversion efficiency of a bulk heterojunction solar cell in dependence of charge carrier mobility equal for electrons and holes, for different surface recombination velocities (solid lines). Under the assumption of a capped Langevin recombination (vertical lines at $10^{-8}$ and $10^{-4}$m$^2$/Vs), a finite minority surface recombination enhances the solar cell efficiency to a saturated level at higher mobilities (dashed lines). Finite majority surface recombination velocities generally lead to a reduction of the power conversion efficiency.} 
	\label{fig:mob}
\end{figure}

The calculated solar cell power conversion efficiencies for varied charge carrier mobilities, equal for electrons and holes, are plotted in Figure \ref{fig:mob}. The power conversion efficiency has an absolute maximum at a certain mobility (here around $10^{-7}\text{m}^2\text{/Vs}$). Whereas the short circuit current increases with higher mobilities, the rather low power converison efficiencies at high mobilities are due to a drastical decrease of the open circuit voltage as already discussed in
\cite{deibel2008a}. Taking finite minority surface recombination velocities of $10^{-4}$m/s, similar to the model of Scott and Malliaras \cite{scott1999a}, this maximum is shifted to higher efficiencies. Taking the majority surface recombination rate additionally into account, we obtain a reduced power conversion efficiency.\\

\begin{figure}
	\centering\includegraphics[width=9.0cm]{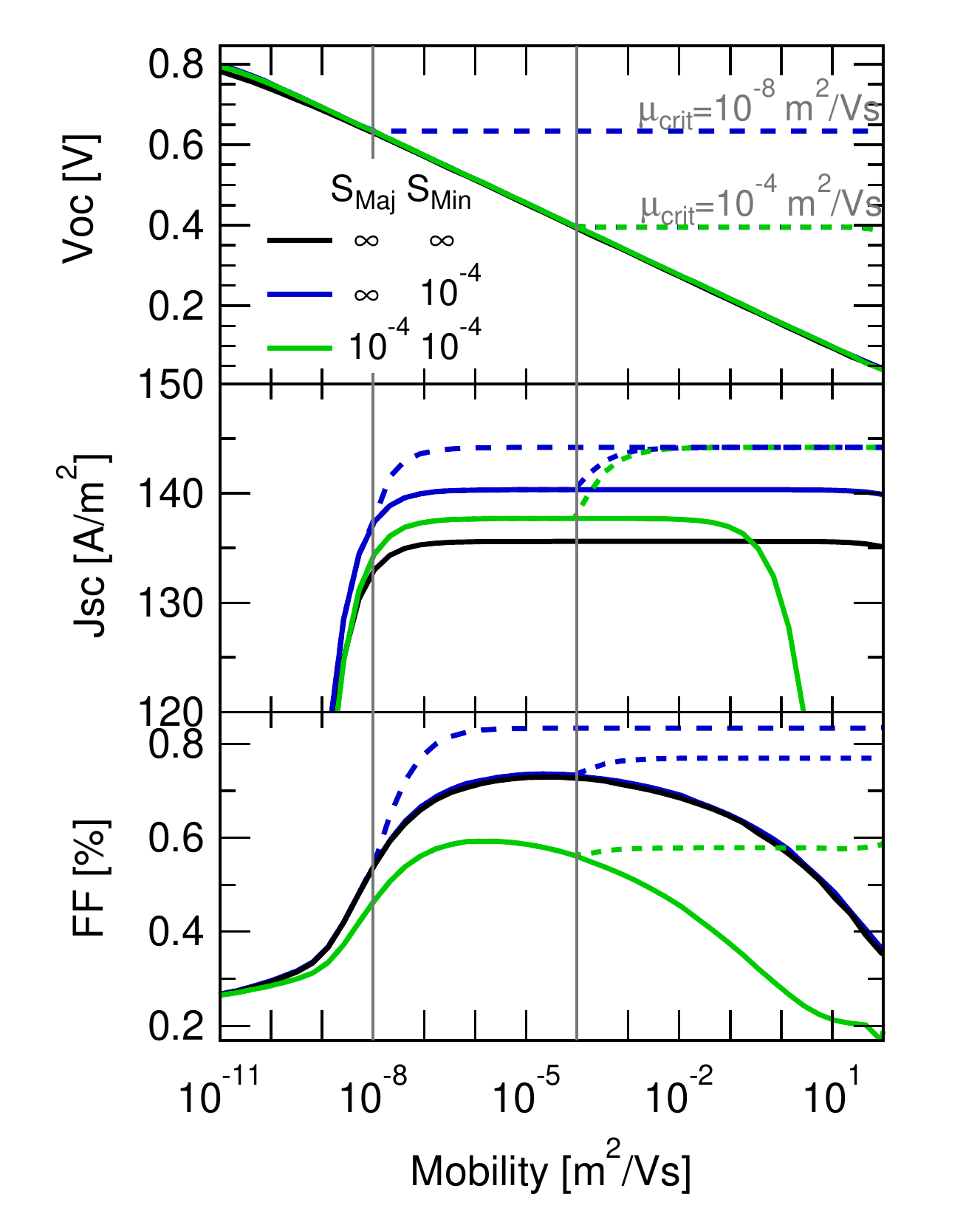}
	\caption{(Color online) Charge carrier mobility dependence of (top) open circuit voltage, (middle) short circuit current density and (bottom) fill factor, resulting in the power conversion efficiencies presented in Figure \ref{fig:mob}. The characterisics are shown for a reduced Langevin recombination (solid, $\zeta = 0.1$) as well as for capped recombinations at mobilities of $\mu_{crit} = 10^{-8}$ and $10^{-4}$m$^2$/Vs for different surface recombination velocity combinations. In general, only the combination of a finite minority surface recombination combined with a capped Langevin recombination rate shows an increase of $V_{oc}$, $J_{sc}$ and fill factor at mobilities higher than the critical mobility.}
	\label{fig:mobdetail}
\end{figure}

Introducing capped Langevin recombination rates in accordance with Eqn. (\ref{eqn:cutoff}), $\mu_{crit}$ set to $10^{-8}$ and $10^{-4}$m$^2$/Vs, the power conversion efficiencies start to further increase with mobilities above the critical mobility, leading to a constant and saturated efficiency at higher mobilities. Without finite minority surface recombination velocity, the power conversion efficiency is not influenced by the capped bimolecular recombination. The reason becomes visible by considering the dependencies of $V_{oc}$, $J_{sc}$ and $FF$ on the charge carrier mobility (Fig. \ref{fig:mobdetail}).\\

The open circuit voltage shows a linear dependence on a logarithmic mobility axis, dropping from 0.8V at low mobilities (and low internal electrical fields) linearly to zero for high mobilities (and high internal electrical fields). Obviously, the difference between $E_g$ and $V_{oc}$ is the effective voltage drop due to the electric field inside the device.

A combination of capped Langevin recombination---raising the internal charge carrier densities---and a finite minority charge carrier recombination---preventing charges to be extracted at the wrong contact---result in a higher and saturated $V_{oc}$ at mobilities above $\mu_{crit}$. Without finite minority surface recombination, minority charges are extracted at both electrodes, leading to a reduced minority charge carrier density there. This effect has a strong effect on the band bending, enhancing the internal electric field. Thus, the open circuit voltage decreases.\\

Beginning at low mobilities, the short circuit current rapidly increases to a saturated level for moderate and high mobilities. Under the influence of a finite minority surface recombination rate, this saturated current is further increased due to surface passivation, preventing minority charges to cross the electrode interface so that they finally recombine there with majority charge carriers. Introducing an additional majority surface recombination rate, creating a space charge at the electrodes, $J_{sc}$ is reduced again by enhanced bulk recombination processes in those space charge areas. Furthermore, at high mobilities, s-shaped current--voltage characteristics  reduce the short circuit current. A more detailed view on this effects will be discussed elsewhere \cite{wagenpfahl2010}. Enabling capped bimolecular recombination rates, once again only the combination with an infinite minority surface recombination rate leads to a fast increase of short circuit current to a new saturated value at higher mobilities. Otherwise the short circuit current remains at a constant level or decreases.\\

An analog behavior holds for the fill factor, showing a maximum around a mobility of $10^{-7}$ to $10^{-4}$ m$^2$/Vs. Exeeding the critical mobility suddenly increases the normally hill-shaped fill factor towards a saturation at higher mobilities, under the presence of a reduced minority surface recombination.\\

\subsection{Variation of surface recombination velocities}

\begin{figure}
	\centering\includegraphics[width=9.0cm]{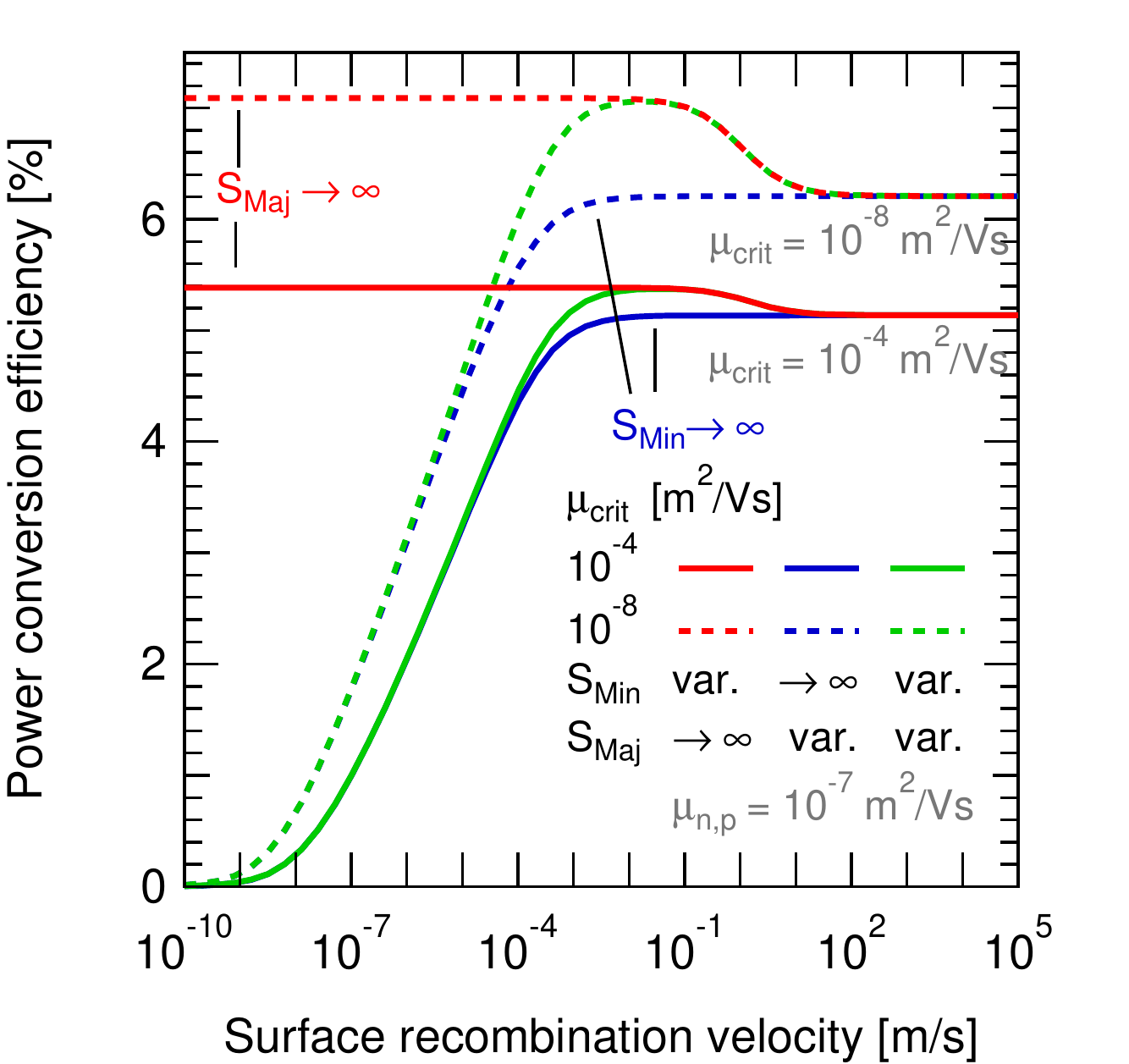}
	\caption{(Color online) Intersection of Figure \ref{fig:mob} at a charge carrier mobility of $10^{-7}$m$^2$/Vs, showing the influence of surface recombination velocities. A decreasing surface recombination first improves the solar cell power conversion efficiency by a dominant minority surface recombination (if existent). At lower values, a possible majority surface recombination reduces the efficiency drastrically. The behaviour for different Langevin capped mobilities remains identical.}
	\label{fig:surf}
\end{figure}

At a charge carrier mobility of $10^{-7}$m$^2$/Vs, the power conversion efficiencies are maximal (Figure \ref{fig:mob}, dotted red line). For this constant mobility, the dependence of the power conversion efficiencies on recombination velocities is shown in Figure \ref{fig:surf}.\\

Decreasing surface recombination velocities starting at infinity down to about $10^1$m/s do not influence the power conversion efficiency at all. In the following range down to $10^{-2}$m/s a dominant minority surface recombination velocity results in an increase of power conversion efficiency until saturation is reached. For velocities lower than $10^{-2}$m/s, the majority surface recombination velocity starts to create space charges in the device leading to a drastic drop of power conversion efficiency. The behaviour of a capped Langevin recombination rate before and after the examined mobility of $10^{-7}$m$^2$/Vs is similar (dotted and solid lines).\\

\begin{figure}
	\centering\includegraphics[width=9.0cm]{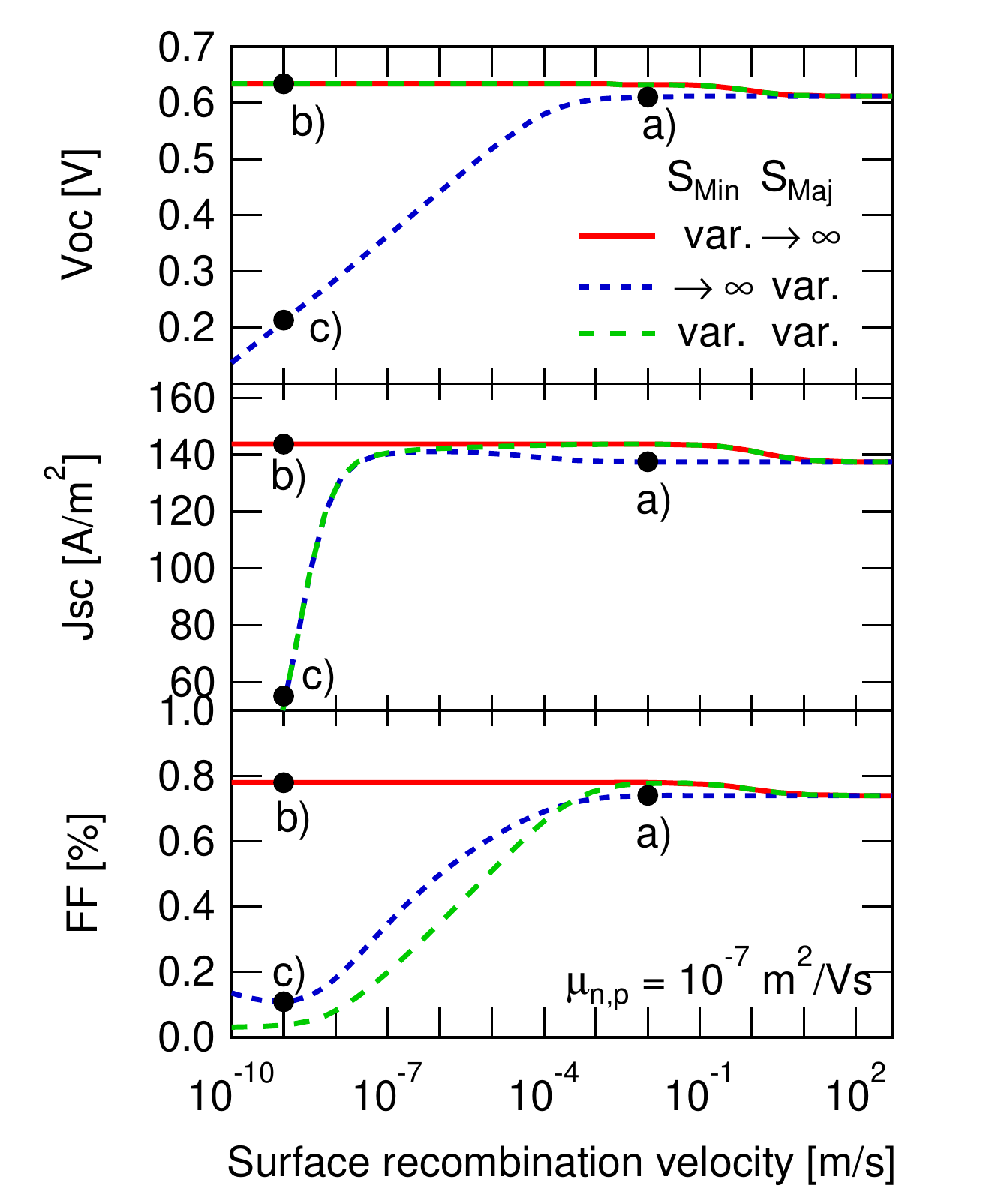}
	\caption{(Color online) Surface recombination dependence of (top) open circuit voltage, (middle) short circuit current density and (bottom) fill factor at different surface recombination velocities at a capped mobility of $10^{-8}$m$^2$/Vs. With decreasing majority surface recombination, $V_{oc}$ as well as $J_{sc}$ decrease. In contrast, a finite minority surface recombination stabilizes $V_{oc}$, the short circuit current density being dominated by majority surface recombination. The fill factor is also dominated by low majority surface recombination velocities.}
	\label{fig:surfdetail}
\end{figure}

\begin{figure}
	\centering\includegraphics[width=9.0cm]{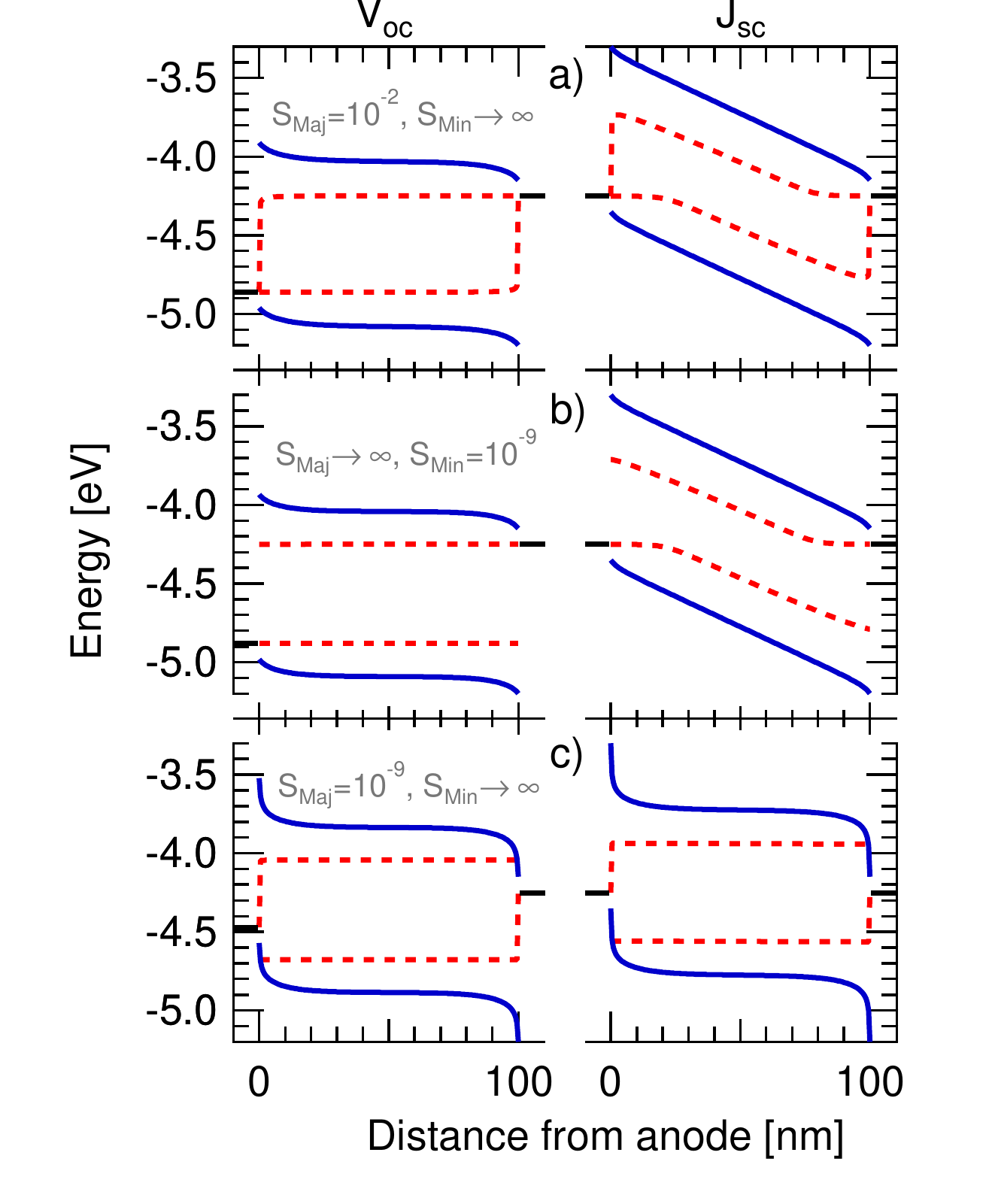}
	\caption{(Color online) Device energy structures at $V_{oc}$ (left column) and $J_{sc}$ (right column) for conditions marked in Figure \ref{fig:surfdetail}. Solid lines indicate the semiconductor LUMO (top) and HOMO (bottom) bands with the corresponding dotted quasi-Fermi levels for electrons (middle up) and holes (middle down). A noninfinite surface recombination velocity introduces a jump of quasi-Fermi levels from the electrodes work function to the semiconductor quasi-Fermi levels.}
	\label{fig:surfband}
\end{figure}

For a critical mobility of $10^{-8}$m$^2$/Vs, $V_{oc}$, $J_{sc}$ and fill factor are shown in dependence of the surface recombination velocity (Figure \ref{fig:surfdetail}). At the marked locations, the corresponding band structures for $V_{oc}$ and $J_{sc}$ conditions are shown in Figure \ref{fig:surfband}. From these band structures the general influence of surface recombination velocities can nicely be seen as infinite values leading to an alignment of the quasi-Fermi levels and the electrodes work functions (Figure \ref{fig:surfdetail} a). In contrast, finite velocities introduce a discontinuity of quasi-Fermi levels at the electrode interfaces as seen in (b) for minorities and (c) for majorities. The loss of minority charge carriers at interfaces can be identified by a high gradient of the quasi-Fermi levels, indicating a (recombination) current towards the electrode ($J_n\propto n(x) \cdot \partial E_{Fn}(x) /\partial x$).\\

The constantly improved power conversion efficiency by finite minority surface recombination velocities (Figure \ref{fig:surf}) can be traced back to a constant improvement of $V_{oc}$, $J_{sc}$ and fill factor (Figure \ref{fig:surfdetail}) as soon as no essential minority recombination current is lost at the contacts (here around $S_{Min}=10^{-1}$m/s). With a finite minority surface recombination, but without any perturbation by majority recombination, the device possesses maximal performance. This is shown in Figure \ref{fig:surfband}b). In contrast, a minority surface recombination current reduces the power conversion efficiency of a device without passivated contacts ( Figure \ref{fig:surfband}a).\\

Starting from $S_{Maj}=10^{-3}$m/s, lower finite majority recombination velocities degrade the solar cell efficiency. As can be seen in Figure \ref{fig:surfband} c), the quasi-Fermi levels of the majority charge carrieres at the interfaces possess a discontinuity to the work function. Accordingly, charge carriers pile up inside the device, which increases the probability to recombine. Therefore, $V_{oc}$ starts to decrease linearly on a logarithmic scale (Figure \ref{fig:surfdetail}). Also the fill factor is reduced. At velocities smaller than $S_{Maj}=10^{-7}$m/s, the short circuit current drops drastically.\\

Bringing both surface recombination effects together (Figure \ref{fig:surfdetail}, green line) leads to a constant and enhanced open circuit potential which is dominated by $S_{Min}$. The short circuit current as well as the fill factor first benefit from the positive influences of the minority surface recombination, but then are decreased by a dominant majority surface recombination, being even more pronounced in case of the fill factor. 

\section{Conclusion}

A lot of research efforts have been put into investigating the bulk recombination in bulk heterojunction solar cells. Up to now, the effect of surface recombination has been largely ignored. Nevertheless, both types of recombination need to be considered as they are crucial for optimizing the power conversion efficiency of organic photovoltaic devices. Previously, it has been shown that organic solar cells have an efficiency peak atmoderatemobilities, but the efficiency decreases again, driven by a lowered open circuit voltage, at higher mobilities. In this contribution, we have shown that this behavior can only be observed for perfectly conducting contacts, implying ideal charge extraction and Langevin recombination, in which the decay rate is limited by the probability of charges finding one another for recombination. However, both assumptions are not generally applicable. In organic devices, the charge extraction is not ideal, and Langevin recombination is only reasonable for rather low mobilities; for fast charge carriers, the recombination rate limiting step is not the finding, but the actual recombination process. Therefore, in our calculations of power conversion efficiencies for bulk heterojunction solar cells, we considered a bulk recombination following Langevin behaviour at low mobilities and a constant rate beyond. Furthermore, we accounted for surface passivating finite minority surface recombination velocities. The combination of both physically justified processes leads to enhanced power conversion efficiencies which do not decrease once a maximum for a certain mobility has been reached, but increase further and finally saturate at high mobilities. Our results thus show that a mobility optimisation can yield more in terms of performance then previously anticipated.

\section*{Acknowledgment}
This work was supported by the German Bundesministerium f{\"u}r Bildung und Forschung, project Massengedruckte Organische Papier-Solarzellen under Contract 13N9867. The work of V.Dyakonov at Bavarian Center for Applied Energy Research was supported by the Bavarian Ministry of Economic Affairs, Infrastructure, Transport and Technology.


\end{document}